\documentclass[12pt]{article}
\usepackage{amsfonts}
\usepackage{amssymb}
\usepackage{color}
\usepackage{graphicx}

\newcommand{\E}{{\cal E}}
\newcommand{\F}{{\cal F}}
\newcommand{\Sc}{{\cal S}}

\newcommand{\mc}{\mathcal}

\newcommand{\be}{\begin{equation}}
\newcommand{\en}{\end{equation}}
\newcommand{\bea}{\begin{eqnarray}}
\newcommand{\ena}{\end{eqnarray}}
\newcommand{\beano}{\begin{eqnarray*}}
\newcommand{\enano}{\end{eqnarray*}}

\newcommand{\A}{{\cal A}}

\newcommand{\Ne}{\overline{\E}}

\renewcommand{\l}{\langle}
\renewcommand{\r}{\rangle}
\newcommand{\pin}[2]{\l#1 , #2\r}

\newcommand{\1}{1 \!\! 1}

\newcommand{\Hil}{\mc H}

\catcode `\@=11 \@addtoreset{equation}{section}

\catcode `\@=12
\begin{document}

\begin{center}
{\Large \textbf{A {\em raised hand effect} as a decision making process}} \vspace{2cm%
}\\[0pt]

{\large F. Bagarello}
\vspace{3mm}\\[0pt]
Dipartimento di Ingegneria,\\[0pt]
Universit\`{a} di Palermo, I - 90128 Palermo,\\
and I.N.F.N., Sezione di Catania\\
E-mail: fabio.bagarello@unipa.it\\

\vspace{7mm}

\end{center}

\vspace*{2cm}

\begin{abstract}
\noindent In this paper we will analyse a group of agents and their attitude to follow, or not, some rules. The model is based on some quantum-like ideas, and in particular on an Hamiltonian operator $H$ describing the dynamics of the agents, assuming they are driven by some mutual interactions and that they are subjected to an external source of "information" used by the agents to decide whether to obey or not these rules. We will discuss how the relative strengths of the  parameters of $H$ determine this attitude and we will discuss in particular the role of the external information. We will also apply our general idea to a specific situation, involving drivers and pedestrians trying to cross a road.
\end{abstract}

\vspace*{1cm}


\vfill

\newpage


\section{Introduction}

Decision Making is a very interdisciplinary topic. It involves, among others, psychological,  mathematical and physiological aspects, and has applications in a huge number of research fields, from Finance to Engineering, from Physics to Biology. We refer to \cite{baa}-\cite{buse3} for some monographs and papers on this topic, where many other references can also be found.

We should stress, before going on, that in all these references the common idea is in using quantum ideas in contexts which are outside what is usually considered {Quantum Mechanics}. The reason for that is related to some specific applications which are more naturally described in terms of operators, rather than of functions. Another reason for this choice is more pragmatic: there are many macroscopic effects which cannot be easily explained using {\em ordinary tools} like differential equations, Newton's laws or stochastic methods. For this reason, we believe, it makes sense to look for other possibilities, and indeed quantum tools and quantum ideas, with Hilbert spaces, non commuting operators, quantum dynamics and so on, have recently proved to provide an interesting alternative possibility.

With this in mind, let us now describe what really inspired the analysis described in this paper: in a crowded city it is quite easy to notice that the uneducated car driver simply does not care much about rules. In particular, the pedestrian crossings are not really taken too seriously by many drivers. Palermo, my city, is one of these crowded cities (not the only one!) where crossing a road could be not so easy. However, a simple solution seems to exist. My personal experience, shared with many other pedestrians, is that the uneducated car driver is subjected to an internal change, becoming very kind and educated,  when the walking person trying to cross the road rise her/his hand, just to ask a bit of attention to the driver: {\em Hello! I am here. I am trying to cross the road. Please, stop your car!} This is what we call {\em the raised hand effect} (RHE). It seems that the simple act of rising an hand creates a deep connection with the driver, who now looks at you not as somebody who is making him late to some meeting, but as someone who needs to be protected! If we make this concrete situation more general and abstract, the idea we are interested here is to describe a situation in which a group of people, the {\em agents}, oscillates between two different attitudes: they do obey some rules, or they don't. The rule, for our concrete drivers/pedestrians system, is clearly {\em don't kill the pedestrian}. In what follows we will propose a simple model to describe this oscillation, and in particular we will show that, within the limit of our model, it exists a concrete parameter or, even better, a specific interplay between some parameters of the model, driving this transition. Our approach is based on similar results deduced recently in \cite{ptrsa}, and uses a particular class of states, the KMS states, which extend the Gibbs states to an algebraic settings and for potentially infinite-dimensional systems. KMS states are used in statistical mechanics to find conditions (if any) for which a specific physical system undergoes a phase transition, \cite{brattrob1,brattrob2}, and for which some equilibrium of the physical system is possible. In \cite{ptrsa} the same class of states has been used to analyze a similar problem in a decision making context, where a group of people was asked to choose between two opposite answers to a certain question. In that situation, it was observed that our model shares some common features with what is called {\em social laser} in \cite{andrei3}, where a small effect produces a big consequence.

 In what we are going to discuss in this paper the situation is slightly different: no {\em phase transition} or {\em social lasing} is really observed, in a statistical mechanics sense, \cite{ruelle}. What we will find, however, is not so unlike: for certain parameters of the model we are going to propose in Section \ref{sect3} there exists a solution which corresponds to having a mostly educated group of agents: the majority of the agents does follow the rules. For other values of the parameters, they do not! So the main role in respecting or not the rules (and respecting or not the pedestrian) is based on a relation between two parameters of the Hamiltonian, $J$ and $B$ (see Section \ref{sect3}), as we will conclude at the end of our analysis.

The paper is organized as follows: in the next section we introduce the essential ingredients of the mathematical settings needed in our analysis. More details on this settings can be found, for instance, in \cite{bagbook2,brattrob2,sewbook1,sewbook2}. In Section \ref{sect3} we describe the model and we analyze the results, discussing, in particular, the relation brtween $J$ and $B$ for the existence of an educated set of agents. Section \ref{sect4} contains our conclusions and plans for the future.

\section{The mathematical settings}\label{sect2}

Let $\Sc$ be a group of $N$ agents who can behave in two opposite ways. Each agent can "obey the rules" or not. We say that the first is an educated agent, while those who do not obey the rules are uneducated, and we indicate with $\E$ those who obey the rules, and with $\Ne$ the others. Hence, it is natural to imagine that each agent $\tau_j$, $j=1,2,\ldots,N$, is described by a linear combination of two orthogonal vectors, one representing $\E$, $e_j^+=\left(
\begin{array}{c}
	1 \\
	0 \\
\end{array}
\right)$, and a second vector, $e_j^-=\left(
\begin{array}{c}
	0 \\
	1 \\
\end{array}
\right)$, corresponding to $\Ne$. These two vectors  form an orthonormal (o.n.) basis in the Hilbert space $\Hil_j=\mathbb{C}^2$, endowed with its standard scalar product $\pin{.}{.}_j$. A general vector describing $\tau_j$ can therefore be written as
\be
\psi_j=a_j  e_j^++b_j e_j^-,
\label{21}\en
with $|a_j|^2+|b_j|^2=1$, to ensure the normalization of the vector $\psi_j$,  $j=1,2,\ldots,N$. The state $\psi_j$ describes a  somewhat intermediate situation, in which $\tau_j$ is potentially both educated and uneducated: $\tau_j$ is in a superposition of states. We introduce the Pauli matrices
$$
\sigma_j^3=\left(
\begin{array}{cc}
	1 & 0 \\
	0 & -1\\
\end{array}
\right), \qquad \sigma_j^+=\left(
\begin{array}{cc}
	0 & 1 \\
	0 & 0\\
\end{array}
\right), \qquad \sigma_j^-=\left(
\begin{array}{cc}
	0 & 0 \\
	1 & 0\\
\end{array}
\right). 
$$
For different $j$, the vectors $e_j^\alpha$ and the matrices $\sigma_j^\beta$, $\alpha=\pm$ and $\beta=3,\pm$, are all copies of the same vectors and matrices. Different $j$ clearly corresponds to different agents of $\Sc$. 

It is clear that 
\be
\sigma_j^3e_j^{\pm}=\pm e_j^{\pm}, \qquad \sigma_j^\pm e_j^{\mp}=e_j^{\pm}, \qquad \sigma_j^\pm e_j^{\pm}=0.
\label{21b}\en
Because of these formulas, we can consider $\sigma_j^3$ as an operator {\em measuring} the decision of $\tau_j$, according to its eigenvalue: if $\tau_j$ belongs to $\E$, then it is described by $e_j^+$, which is an eigenstate of $\sigma_j^3$ with eigenvalue $+1$. Similarly, if $\tau_j$ belongs to $\Ne$, then it is described by $e_j^-$, which is also an eigenstate of $\sigma_j^3$, but with eigenvalue $-1$. Of course, $\sigma_j^\pm$ are operators which modify the original attitude of $\tau_j$, according to (\ref{21b}). For instance, the action of $\sigma_j^+$ on the agent $\tau_j$ originally in $\Ne$ moves the agent to $\E$. However, if {\em somebody} insists too much, then it is like we are acting more than once with $\sigma_j^+$ (or with $\sigma_j^-$), and the result is that $\tau_j$ is moved to the zero vector! The agents don't like to be stressed too much.

\vspace{2mm}

{\bf Remark:--} We could replace the Pauli matrices with two-by-two matrices obeying the canonical anti-commutation relations, \cite{mer,bagbook}. Indeed $\sigma_j^+$ and $\sigma_j^-$ respectively behave like two well known  fermionic creation and annihilation operators $c$ and $c^\dagger$, while $\sigma_j^3$ is the related number operator, $\hat n=c^\dagger c$, which should be identified (to respect the eigenvalues) with $\frac{1}{2}(\1-\sigma_j^3)$.

\vspace{2mm}

Out of $\sigma_j^\pm$ we can also introduce the other two Hermitian Pauli matrices
$$
\sigma_j^1=\sigma_j^++\sigma_j^-=\left(
\begin{array}{cc}
	0 & 1 \\
	1 & 0\\
\end{array}
\right),\qquad \sigma_j^2=i(\sigma_j^--\sigma_j^+)=\left(
\begin{array}{cc}
	0 & -i \\
	i & 0\\
\end{array}
\right), 
$$
and the following rules are easily found:
\be
(\sigma_j^\pm)^2=0, \qquad [\sigma_j^\pm,\sigma_j^3]=\mp \,2\sigma_j^\pm\qquad [\sigma_j^+,\sigma_j^-]=\sigma_j^3,\qquad \{\sigma_j^+,\sigma_j^-\}=\1,
\label{21c}\en
where $[A,B]=AB-BA$ and $\{A,B\}=AB+BA$ are the commutator and the anti-commutator of the operators $A$ and $B$.

The Hilbert space for $\Sc$ is made of copies of $\mathbb{C}^2$, one for each agent:
$$
\Hil=\otimes_{j=1}^{N}\mathbb{C}_j^2.
$$

An o.n. basis of $\Hil$  consists of tensor products of states $e_j^+$, and $e_j^-$ for various $j$. For instance, the vector describing a situation in which the first $L_1$  agents of $\Sc$ are in $\E$ and the remaining $L_2$ agents belong to $\Ne$, $L_1+L_2=N$,  is 
\be
\Phi=(e_1^+\otimes\cdots\otimes e_{L_1}^+)\otimes(e_1^-\otimes\cdots\otimes e_{L_2}^-).
\label{21d}\en
Of course the dimensionality of $\Hil$ increases with $N$. In fact we have $\dim(\Hil)=2^N$. An operator $X_1$ acting, for instance, on $\mathbb{C}_1^2$, is identified with the tensor product $X_1\otimes \1_2\otimes\cdots\1_N$, i.e. the tensor product of $X_1$ with $N-1$ copies of the identity operator $\1$, acting on all the other single-agent Hilbert spaces. In this way we can relate $X_1$ with a (bounded) operator on $\Hil$. The set of all the linear (bounded) operators on $\Hil$ defines an algebra $\A$, which we can call, adopting a standard nomenclature, {\em the algebra of the observables}.
The vector $\Phi$ above, and all other normalized vectors on $\Hil$, defines a vector state on $\A$, \cite{brattrob1,sewbook1}, as  follows: $\omega_\Phi(A)=\pin{\Phi}{A\,\Phi}$, $A\in\A$. Here $\pin{.}{.}$ is the scalar product on $\Hil$,
$$
\pin{f_1\otimes\cdots\otimes f_N}{g_1\otimes\cdots\otimes g_N}=\prod_{j=1}^{N}\pin{f_j}{g_j}_j,
$$
$f_j,g_j\in\mathbb{C}_j$.
Vector states are only one class among all the possible states over $\A$. Other examples of states, i.e. (from a mathematical point of view) of positive normalized functionals on $\A$, are Gibbs and KMS states, \cite{brattrob2,sewbook1}. These are particularly important because of their physical interpretation in statistical mechanics: they are both equilibrium states at a non zero temperature, while $\omega_\Phi$ above typically describe a zero-temperature situation. The difference between Gibbs and KMS states is that, usually, Gibbs states are used for finite dimensional systems, while KMS states are their counterpart for systems with an infinite number of degrees of freedom. Both Gibbs and KMS states satisfy the so-called {KMS condition}, which we write here in a simplified version. We refer to \cite{brattrob2,sewbook1,sewbook2} for more details, and for a deeper mathematical analysis, of KMS states. The KMS condition is
\be
\omega(AB)=\omega(BA(i\beta)),
\label{22}\en
where $A,B\in\A$, $A(t)$ is the time evolution of $A$ and $A(i\beta)$ is the analytic continuation of $A(t)$ computed in $t=i\beta$, $\beta=\frac{1}{T}$, the inverse temperature (in units in which the Boltzmann constant is equal to one). In the Heisenberg picture, if $H$ is the time-independent Hamiltonian of the system, $A(t)$ is the following operator: $A(t)=e^{iHt}Ae^{-iHt}$. Also, $\omega$ can be written as
\be
\omega(A)=tr(\rho A), \qquad \rho=\frac{e^{-\beta H}}{tr(e^{\beta H})}.
\label{23}\en
Here $tr(X)$ is the trace of the operator $X$, and the KMS condition can be deduced (at least formally) using the cyclicity of the trace, together with the explicit expressions of $\rho$ and $A(t)$.

\vspace{2mm}

{\bf Remark:--} More explicitly, given any o.n. basis $\F_c=\{c_n\}$,  a
Gibbs state can be rewritten as follows: $$
\omega(X)=\frac{1}{Z}\sum_n\left<c_n,{\sf e}^{-\beta
	H}Xc_n\right>, $$ where $Z=\sum_n\left<c_n,{\sf
	e}^{-\beta H}c_n\right>$ {and $\beta$ is the inverse
	temperature, always positive}. This expression for $\omega$ is equivalent to the one adopted in (\ref{23}). Despite of its apparently complicated expression $\omega(X)$ can be often computed in a reasonably easy way choosing, for $\F_c$, the set of the eigenstates of $H$, if they can be found. This is because the trace of an operator is independent of the o.n. basis used to compute it, \cite{rs}.  

\vspace{2mm}

  We refer to \cite{ptrsa} for more results which are interesting for us, and to \cite{brattrob2,sewbook1,sewbook2} for more details on the so-called algebraic dynamics for quantum systems, which plays a crucial role in this context. As we have already observed, for us here the equality in (\ref{22}) is particularly relevant since, as in statistical mechanics, it produces conditions for the system to admit some equilibrium.

\section{The model}\label{sect3}

The system $\Sc$ we want to describe is made by $N$ agents, which we could imagine to be drivers in a crowded city (like Palermo), moving around and respecting, or not, the rules of the streets. A state $\Phi_\uparrow=\otimes_{j=1}^Ne_j^+$ describes a set of agents all respecting the rules, while $\Phi_\downarrow=\otimes_{j=1}^Ne_j^-$ describes, on the other hand, a set of agents {\bf not} respecting the rules. We observe that, if we introduce the operator $\sigma_N^3=\frac{1}{N}\sum_{i=1}^{N}\sigma_i^3$ and its mean values on a certain (product) state $\Phi$, see (\ref{21d}) as an example, it turns out that, in particular, 
$$
\lim_{N,\infty}\Omega_\uparrow(\sigma_N^3):=\lim_{N,\infty}\langle\Phi_\uparrow, \sigma_N^3\Phi_\uparrow\rangle=1, \qquad\quad \lim_{N,\infty}\Omega_\downarrow(\sigma_N^3):=\lim_{N,\infty}\langle\Phi_\downarrow, \sigma_N^3\Phi_\downarrow\rangle=-1.
$$
This suggests to associate the (limit for $N\rightarrow\infty$ of the) mean value of $\sigma_N^3$ on a generic state on $\Hil$, not necessarily of the vector form\footnote{We recall that these are the states of the form $\langle\Phi, \cdot\,\Phi\rangle$, for some normalized $\Phi\in\Hil$.} , with a sort of {\em order parameter} describing the tendency of the agents to obey the rules (when this value is +1) or not (when its value is -1). In other words, if $\omega$ is a state of the system, if $\lim_{N,\infty}\omega(\sigma_N^3)>0$, then the agents are mostly respecting the rules, and the more this limit approaches $+1$, the more they do. On the other hand,  if $\lim_{N,\infty}\omega(\sigma_N^3)<0$, then the agents are mostly neglecting the rules, and the more this limit approaches $-1$, the more they are uneducated. If $\lim_{N,\infty}\omega(\sigma_N^3)=0$ there are agents respecting the rules, and (almost) the same number of agents not respecting the rules. 

This attitude of the agents, in our model, is the result of the interactions among themselves, and of their simultaneous interaction with some {\em external  source}. In this perspective, the system is, in a certain sense, {\em open}. As we will see in what follows, the main ingredients of our model here are similar to those considered in \cite{ptrsa}, even if some relevant differences are necessarily introduced, and these differences contribute to make our present model interesting.

According to what we have seen in Section \ref{sect2}, the single agent is described in terms of Pauli matrices: in particular, we have {\em changing attitude} operators, $\sigma_j^\pm$, and the {\em checking attitude} operator $\sigma_j^3$. The full system $\Sc$ is made by $N$ agents, and the description lives in $\Hil=\otimes_{j=1}^{N}\mathbb{C}_j^2$, as discussed above. As often in these cases, \cite{bagbook,bagbook2}, the dynamics of $\Sc$ is given in terms of an Hamiltonian operator $H_N$ which is constructed in order to describe the (hopefully) most relevant effects occurring in $\Sc$. The first effect we want to consider is described by the following term:
\be
H_1^{(N)}=\sum_{i,j=1}^{N}\,J_{i,j}\sigma_i^3\sigma_j^3.
\label{31}\en 
As in \cite{ptrsa}, we observe that if the {\em two-body potential} $J_{i,j}$ is positive, for all $i,j=1,2,\ldots,N$, then the contribution to the eigenvalues of $H_{1}^{(N)}$ is positive if $\tau_i$ and $\tau_j$ are both in the up vector, $e_i^+$ and $e_j^+$, or both in the down vector, $e_i^-$ and $e_j^-$. On the other hand, if $J_{i,j}<0$, then the contribution of $J_{i,j}\sigma_i^3\sigma_j^3$ to the eigenvalues of $H_{1}^{(N)}$ is positive only when $\tau_i$ and $\tau_j$ are in opposite vector,  $e_i^+$ and $e_j^-$, or  $e_i^-$ and $e_j^+$, and negative if $\tau_i$ and $\tau_j$ are described by the same vector (both up, or both down). Hence the (energetic) effect of $H_{1}^{(N)}$ is deeply connected with the sign of $J_{i,j}$. In this paper, to fix the ideas, we will only consider the case of negative $J_{i,j}$: $J_{i,j}<0$, $\forall i,j$. Hence the effect of $H_1^{(N)}$ would be to force all the agents to behave similarly (all up, or all down).

As in \cite{ptrsa}, we also consider the following operator: 
\be
H_{2}^{(N)}=\sum_{i,j=1}^{N}\,p_{i,j}\left(\sigma_i^+\sigma_j^-+\sigma_i^-\sigma_j^+\right),
\label{32}\en
whose effect is more psychological than energetic: it describes a situation in which, during an interaction between $\tau_i$ and $\tau_j$, the two agents tend to act in an opposite way: if $\tau_i$ changes from a "obey" to a "not obey" behavior (because of $\sigma_i^-$), $\tau_j$ moves in the opposite direction, because of $\sigma_j^+$. Hence, $H_{2}^{(N)}$ describes an effect which is somewhat opposite with respect to that produced by (\ref{31}) where, since $J_{i,j}<0$, the energetically preferred vectors were $\Phi_\uparrow$ and $\Phi_\downarrow$, both with all parallel spins. Of course, $p_{i,j}$ gives the {\em strength} of the mutual tendency of the agents to behave differently.

A third term in the Hamiltonian of $\Sc$ is the following {\em one-body} term:
\be
H_{3}^{(N)}=B\sum_{i=1}^{N}\,\sigma_i^3=BN\sigma_N^3,
\label{34}\en
$B\in\mathbb{R}$, which was only mentioned in \cite{ptrsa} but not considered in the dynamics of the system analyzed there. Here, on the other hand, this term is one of the essential ingredients of our model, since $B$ is, as it will be explained soon, one of the parameters which drive the agents towards an {\em ordered} or a {\em disordered} state. Here ordered or disordered should be interpreted as a state $\Omega(.)$ which produces respectively a positive or a negative value of $\lim_{N,\infty}\Omega(\sigma_N^3)$, since this is what gives, as we have already discussed, a measure of the attitude of the agents to obey or not the rules. Hence, ordered and disordered have nothing to do with the entropy, also in view of the fact that $\Omega_\uparrow$ and $\Omega_\downarrow$ are both rather ordered (from an entropy point of view), but describe completely different attitudes of the agents in the context of our model. In a certain sense, rather than ordered or disordered, we could use educated or uneducated states, which is a nomenclature related to what we are describing, but also quite {\em strange}\footnote{This is the reason why we prefer the other adjectives for the states, ordered and disordered.}. 

 We use an energetic understanding of $H_{3}^{(N)}$, as we did for $H_1^{(N)}$: if $B>0$, then the lowest eigenvalue of $H_{3}^{(N)}$ is obtained by a vector with all down single agent vectors, $\Phi_\downarrow=e_1^-\otimes\cdots\otimes e_{N}^-$, and its corresponding eigenvalue is clearly $-NB$. Analogously, if we take $B<0$, then the lowest eigenvalue of $H_{3}^{(N)}$ is $NB$, with corresponding eigenvector  $\Phi_\uparrow=e_1^+\otimes\cdots\otimes e_{N}^+$. It is clear then that  $H_{3}^{(N)}$ can be used to model an external source (which here we consider as a sort of information, e.g. the rising hand!) driving $\Sc$ toward different states, depending on the sign of $B$. It might be useful to observe that this is a different effect with respect to the one described by $H_1^{(N)}$, where different values of the energy is related to having single agents states parallel or anti-parallel. Here, for $H_3^{(N)}$, the minimum in energy is achieved, both for $B>0$ and for $B<0$, for parallel vectors for the various agents.

Putting all these contributions together, we introduce the Hamiltonian
\be
h_N=H_{1}^{(N)}+H_{2}^{(N)}+H_{3}^{(N)}=\sum_{i,j=1}^{N}\,J_{i,j}\sigma_i^3\sigma_j^3+\sum_{i,j=1}^{N}\,p_{i,j}\left(\sigma_i^+\sigma_j^-+\sigma_i^-\sigma_j^+\right)+B\sigma_N^3,
\label{33}\en
which will be the starting point of our analysis.
\vspace{2mm}
Following similar steps as in \cite{ptrsa}, and in particular adopting the  {\em mean field approximation}:
$$
p_{i,j}\rightarrow \frac{p}{N}, \qquad J_{i,j}\rightarrow \frac{J}{N},
$$
with $p=2J$, we can replace  $h_N$ with its mean field approximation $H_N$:
\be
H_N=\frac{J}{N}\sum_{i,j=1}^{N}\sum_{\alpha=1}^3\sigma_i^\alpha\sigma_j^\alpha+BN\sigma_N^3,
\label{35}\en
which is a mean-field Heisenberg model in presence of a magnetic field, \cite{ruelle}. Due to our working hypothesis on $J_{i,j}$, we have $J=-|J|<0$. Hence $p<0$ as well. The existence of the dynamics for $H_N$, and of the thermodynamical limit of this model ($N\rightarrow\infty$), has been discussed at length in, for instance, \cite{bagmorc,bagmaster}, where it is  proved, among other results, what is listed below:

\begin{enumerate}

\item the operator $\sigma_N^\alpha=\frac{1}{N}\sum_{i=1}^{N}\sigma_i^\alpha$, $\alpha=1,2,3$, converges (in the strong topology restricted to a suitable set of states) to an operator $\sigma_\infty^\alpha$ which commutes with all the $\sigma_k^\beta$, $\forall k$ and $\forall\beta$, and therefore belongs to the center of $\A$.

\item $H_N$ generates a time evolution of each $\sigma_\alpha^i$, $\alpha_N^t(\sigma_\alpha^i)=e^{iH_Nt}\sigma_\alpha^ie^{-iH_Nt}$, which converges (again, in the strong topology), to
\be
\alpha^t(\sigma^\alpha_i)=\cos^2(Ft)\sigma^\alpha_i+\frac{i}{F}\sin(Ft)\cos(Ft)[\underline{F}\cdot\underline\sigma_i,\sigma^\alpha_i]+\frac{1}{F^2}\sin^2(Ft)(\underline{F}\cdot\underline\sigma_i) \,\sigma^\alpha_i (\underline{F}\cdot\underline\sigma_i),
\label{36}\en
	where $\underline{F}=J\underline \sigma_\infty+\underline{B}$, $\underline{B}=(0,0,B)$, and $F=\|\underline{F}\|$. We observe that (\ref{36}) is well defined for all $F\neq0$, and that it can also be extended continuously to the case $\underline F=\underline 0$. In this case, it is easily checked that $\alpha^t(\sigma^\alpha_i)=\sigma^\alpha_i$: the time evolution trivializes. This aspect will appear again in the following.
	
	\item If we are now interested in the expression of $\alpha^t(\sigma^\alpha_i)$ in a representation $\pi_\omega$, arising as the GNS-representation\footnote{The mathematical details of this construction are not very relevant for us, and can be found in \cite{bagmorc}. The relevant aspect here is that different representations correspond to different physics for the same system, \cite{brattrob1,brattrob2}. Here it is sufficient to stress that a representation is a concrete realization of an abstract algebra.} of a state $\omega$ over $\A$, formula (\ref{36}) produces
	\be
	\alpha_\pi^t(\sigma^\alpha_i)=\cos^2(ft)\sigma^\alpha_i+\frac{i}{f}\sin(ft)\cos(ft)[\underline{f}\cdot\underline\sigma_i,\sigma^\alpha_i]+\frac{1}{f^2}\sin^2(ft)(\underline{f}\cdot\underline\sigma_i) \,\sigma^\alpha_i (\underline{f}\cdot\underline\sigma_i),
	\label{37}\en
	where $\underline f=\pi_\omega(\underline F)=J\underline{ m}+\underline B$, where $\underline m=\pi_\omega(\underline \sigma_\infty)$ and $f=\|\underline f\|$. Here and in (\ref{36}) $\alpha=1,2,3$, while $i=1,2,3,\ldots,N$. It might be useful to stress that $\underline m$, and the related vector $\underline f$, are the counterparts of $\underline \sigma_\infty$ in the $\pi_\omega$ representation. This means that they are related to the order parameter which describes the overall tendency of the agents to obey or not the rule. In particular, this is exactly the role of $m_3$. The other components of $\underline m$, $m_1$ and $m_2$ appear (and play a role) because of the non commutativity of our model.
	
	\item The effective dynamics $\alpha_\pi^t$ can be deduced by the following effective, representation-dependent, Hamiltonian:
	$$
	H_\pi=\underline f\cdot\sum_{i=1}^\infty \underline \sigma_i.
	$$
	Of course, using $H_\pi$ rather than $H_N$ simplifies the analysis of the dynamics of $\Sc$ quite a bit, \cite{bagmorc,bagmaster}, and, for this reason, it is quite useful in the dynamical analysis of the model.
	
\end{enumerate}

The analytic expression for $H_\pi$ clarifies that the only relevant case is when $\underline f\neq\underline 0$. Otherwise $H_\pi=0$ and the dynamics trivialize: $\alpha_\pi^t(A)=e^{iH_\pi t}Ae^{-iH_\pi t}=A$, for all $A\in\A$. This is, in representation, what we have already observed in point 2. above, at a pure algebraic level.

Following what we did in \cite{ptrsa} we focus here on the KMS identity (\ref{22}) with formula (\ref{37}) for the dynamics. Due to the properties of the GNS representation, it is known that  $\underline m=\pi_\omega(\underline \sigma_\infty)=\omega(\underline \sigma_i)$, $\forall i$. In particular we see that the dependence of the agent index $i$ disappears. This is because all our agents are {\em equivalent} in the treatment discussed here, as the analytic expression of $H_N$ in (\ref{35}) clearly shows. Hence, after some algebra, using (\ref{21c}) and putting $A=B=\sigma_1^i$, we can rewrite the equality in (\ref{22}), $\omega(\sigma_1^i\sigma_1^i)=\omega(\sigma_1^i\sigma_1^i(i\beta))$, as follows:
\be
f\left(J(m_2^2+m_3^2)+Bm_3\right)\cosh(\beta f)+\left(f^2-Jm_1(Jm_1-iBm_2)\right)\sinh(\beta f)=0,
\label{38}\en
where, as already observed, $f=\|\underline f\|>0$, strictly positive, is the only interesting case. Also, we are only interested in considering the case of non zero $B$, since otherwise the model collapses to that in \cite{ptrsa} and, more important, because of the interpretation of $B$ in this paper, as an external source of information  (the pedestrian's hand raised or not) reaching the agents (the car drivers). Needless to say, $J\neq0$ as well, to avoid to completely remove from our model the effect of (the mean field versions of) $H_{1}^{(N)}$ and $H_{2}^{(N)}$. We recall that we are focusing here on $J<0$. Looking now to (\ref{38}) it is clear that we must have, first of all, $JBm_1m_2=0$. This means that, in view of our constraints, either $m_1=0$ or $m_2=0$, or both. In this paper we are not interested in finding all the possible non trivial solutions of (\ref{38}), but only to show that some solution exists, and it has an interesting interpretation. With this in mind, we will  consider here only  the choice $m_2=0$, trying to understand if, with this choice, it is possible to have a non trivial solution of (\ref{38}), and when (i.e., for which range of parameters). 

If $m_2=0$, recalling that $f^2=J^2(m_1^2+m_2^2)+(Jm_3+B)^2=J^2m_1^2+(Jm_3+B)^2$, formula (\ref{38}) can be rewritten as
\be
(Jm_3+B)\left[m_3f\cosh(\beta f)+(Jm_3+B)\sinh(\beta f)\right]=0,
\label{39}\en
which has a first obvious solution: $m_3=-\frac{B}{J}$. With this choice we have $\underline f=(m_1,0,0)$, where $m_1$ should be chosen different from zero, to exclude the case of $\underline f=\underline 0$. However, this is not an interesting solution for us since, with this form of $\underline f$, we can easily check that $\alpha_\pi^t(\sigma^1_i)=\sigma_i^1$. Hence it is obvious that the KMS condition for $A=B=\sigma_1^i$ is always satisfied: it simply becomes the identity "1=1".

For this reason we next assume that $Jm_3+B\neq0$, so that a solution of (\ref{39}) exists if and only if
$$
\Phi(f)=m_3f\cosh(\beta f)+(Jm_3+B)\sinh(\beta f)
$$
has a nontrivial zero. Indeed, $f=0$ is such that $\Phi(0)=0$, but this is not a relevant solution for us, as we have already noticed. Hence we need to check if it is possible to find $\tilde m_1$ and $\tilde m_3$ such that, calling $\tilde {\underline f}=(\tilde m_1,0,J\tilde m_3+B)$, $\Phi(\tilde{\underline f})=0$. To check if such a solution exists we first recall again that $\Phi(0)=0$. Now, observing that $f$ depends on $m_1$ and $m_3$, after some easy computations we deduce that $\Phi'(0)=m_3$. Hence $\Phi(f)$ is an increasing (resp. decreasing) function in $f=0$ if $m_3>0$ (resp $m_3<0$). Now, if $m_3>0$ and $Jm_3+B>0$, $\Phi(f)$ is always increasing and strictly positive, as one also deduce simply looking at the analytic expression of $\Phi(f)$ above:

{\em if $m_3>0$ and $-|J|m_3+B>0$, there is no solution of (\ref{38}) of the form $\underline f=(m_1,0,Jm_3+B)$.}

Hence  a solution of (\ref{38}) for $m_2=0$ and $m_3>0 $ can exist only if $Jm_3+B=-|J|m_3+B<0$, i.e. if $B<|J|m_3$. This is what indeed happens: we will now show that, under these conditions, $\Phi(f)=0$ for some $\underline f=(0,0,Jm_3+B)$, with $Jm_3+B<0$. In fact, in this case we have $f=|Jm_3+B|=-(Jm_3+B)$, and
$$
\Phi(f)\,\rightarrow\,\Phi_{B,J}(m_3)=|Jm_3+B|\left(m_3\cosh\left(\beta(Jm_3+B)\right)+\sinh\left(\beta(Jm_3+B)\right)\right),
$$
where we have replaced the dependence of $\Phi$ on $f$ with that on $m_3$, and we have added the suffix ${B,J}$ to stress the fact that $\Phi$ also depends on these two parameters of the model. We can further rewrite
\be
\Phi_{B,J}(m_3)=\frac{|Jm_3+B|}{2}\left((m_3+1)e^{\beta(Jm_3+B)}+(m_3-1)e^{-\beta(Jm_3+B)}\right).
\label{310}\en
Since $Jm_3+B<0$ here, and $\beta>0$, it follows that if $|Jm_3+B|$ is sufficiently large, then 
$$
\Phi_{B,J}(m_3)\simeq \frac{|Jm_3+B|}{2}(m_3-1)e^{-\beta(Jm_3+B)},
$$
which is surely negative if $m_3<1$. Then $\Phi_{B,J}(m_3)$, after increasing for positive values (since $m_3>0$), becomes negative  if $|Jm_3+B|$ is large enough. Hence it assumes value zero for a (or some) specific combination of $B$, $J$ and $m_3$. Once more we recall that $J<0$, $0<m_3<1$ and $Jm_3+B<0$. Because of the bounds on $m_3$, and assuming that $J$ is fixed, $|Jm_3+B|$ can become large only if $B$ increases sufficiently. But, recalling that $B<|J|m_3<|J|$, $B$ must increase for negative values. Therefore a non trivial solution of the KMS condition for $\sigma_i^1$ can exist only if $B$ in (\ref{33}) satisfies these constraints. Is the magnetic field the parameter of the model that triggers the possibility of having a non trivial solution of equation (\ref{38}), together with the parameter $J$ and, in particular, if $J$ is fixed, then we need to increase the value of $-B$ to get a solution.

It is interesting to observe that $\beta$ plays no particular role, at least concerning the existence of this specific solution. Of course, it affects the explicit value of the solution, but not the fact that it exists or not. This is only related to the original parameters of the mean field Hamiltonian in (\ref{35}).

It is also interesting to observe that it is impossible, in our situation, to obtain full obedience to the rules, i.e. to have a solution of the KMS condition for $m_3=1$. Indeed, with this choice, we should have $m_1=m_2=0$, $m_3=1$, $\underline f=(0,0,B-|J|)$ with $B<|J|$. Hence from (\ref{310}) we see that
$
\Phi_{B,J}(1)=|J+B|\,e^{\beta(J+B)},
$
which is always positive, for all possible choice of $B$ and $J$: no solution of $\Phi_{B,J}(1)=0$ exists, in this case.

\vspace{2mm}

{\bf Remark:--} The case $m_3<0$ is similar to the one considered here. In this case, we should find conditions for $\Phi(f)$ to assume positive values for some $J,B$ and $m_3$. In fact, this guarantees the existence of a non trivial solution of $\Phi(f)=0$, similarly to what we have deduced above for $m_3>0$.

\section{Conclusions}\label{sect4}

In this paper we have proposed a model, based on a spin Hamiltonian, describing a group of agents subjected to an external input forcing the agents to follow, or not, some specific set of rules. In particular, our specific interest was connected to what we have called  {\em RHE}. We have shown that, independently of other possible effects (like the effect of the inverse temperature $\beta$), it is possible to use $B$ as a sort of {\em raised hand}, since when its value satisfies a certain constraint together with $J$, a solution with positive $m_3$ is possible: agents, on average, follow the rules, and stop at the pedestrian crossing! However we have also deduced that, at least under the limits of our simplifications, it is not possible to reach a complete obedience, since $m_3$ cannot be strictly equal to 1.

It is clear that ours is really a preliminary analysis of the system. Other solutions could exist, as well as other no-go results. Moreover, we should still also fully understand the role of the parameter $\beta$: we have already observed that the explicit solution of the KMS condition depends on $\beta$, but we have also seen that a non trivial solution of the KMS condition can be found independently of the value of $\beta$. What is still to be understood is whether $\beta$ plays a role in the existence of other non trivial solutions of the KMS condition. Also, is our Hamiltonian the only possible, or the best one, to describe the RHE? Why KMS and not other classes of states? Is it possible to refine the model including more effects? These are only few of the questions one could imagine to explore next, and in fact these are part of our future plans.

\section*{Acknowledgements}

The author acknowledges partial financial support from Palermo University and from G.N.F.M. of the INdAM. This work has also been partially supported by the PRIN grant {\em Transport phenomena in low dimensional
	structures: models, simulations and theoretical aspects}, by Project {\em CAESAR} and by Project {\em ICON-Q}.

%
%
%
%
%


\begin{thebibliography}{99}
	
\bibitem{baa} B.E. Baaquie, {\em Quantum finance, path integrals and Hamiltonians for options and interest rates}, Cambridge
University Press, (2004)


\bibitem{bagbook} F. Bagarello, {\em Quantum dynamics for classical systems: with applications of the
	Number operator}, John Wiley $\&$ Sons, Hoboken, (2012)

\bibitem{bagbook2} F. Bagarello, {\em Quantum Concepts in the Social, Ecological and Biological Sciences}, Cambridge University Press, Cambridge, (2019)

\bibitem{FFF} F. Bagarello, F. Gargano, F. Oliveri, {\em Quantum Tools for Macroscopic Systems}, Springer, Synthesis Lectures on Mathematics $\&$ Statistics (2023)



\bibitem{Busemeyer2012} J. R. Busemeyer, P. D. Bruza,  \textit{Quantum
	models of cognition and decision},  Cambridge University Press, Cambridge, (2012)



\bibitem{havkhrebook} E. Haven, A.  Khrennikov, {\em Quantum social science}, Cambridge University Press, (2013)

\bibitem{Khrennikov2010} A. Khrennikov, \textit{Ubiquitous quantum
	structure: from psychology to finance}, Berlin-Heidelberg-New York, Springer, (2010)

	
	
	\bibitem{QuantumCognition} E. M. Pothos, J. R. Busemeyer,   {\em Quantum Cognition}, {Annual Review of Psychology}, \textbf{73}, 749-778, (2022)
	
	\bibitem{buse3}J. R. Busemeyer, Z. Wang, J. T. Townsend, {\em Quantum dynamics of human
		decision making}, J. Math. Psych., {\bf50}, 220-241,  (2006)
	
	\bibitem{ptrsa} F. Bagarello, {\em Phase transitions, KMS-condition and Decision Making},   Phil. Trans. R. Soc. A, {\bf 381}: 20220377 (2023)	
	
	\bibitem{brattrob1} O. Bratteli e D.W. Robinson, {\em Operator algebras and quantum statistical
		mechanics}, vol 1, Springer, Heidelberg,  (1979)
	
	\bibitem{brattrob2} O. Bratteli e D.W. Robinson, {\em Operator algebras and quantum statistical
		mechanics}, vol 2, Springer, Heidelberg,  (1981)
	
	\bibitem{andrei3}  A. Khrennikov,
	{\em Social laser}, Jenny Stanford Publ.,
	Singapore, (2020)
	
	\bibitem{ruelle} D. Ruelle, {\em Statistical mechanics. Rigorous results}, Reprint of the 1989 edition,
World Scientific Publishing Co., Inc., River Edge, NJ; Imperial College Press,
London, (1999)
	
	
	
	
	
	
	
	\bibitem{sewbook1} G.L. Sewell, {\it Quantum Theory of Collective Phenomena}, Oxford University Press,
	Oxford, (1989)
	
	\bibitem{sewbook2} G.L. Sewell, {\it Quantum Mechanics and its Emergent Macrophysics}, Princeton University Press,
	(2002)
	
	
	\bibitem{mer} E. Merzbacher, {\em Quantum Mechanics}, Wiley, New York (1970),
	
	\bibitem{rs}  M. Reed and B. Simon,
	{\it Methods of Modern Mathematical Physics I, Functional Analysis}, Academic Press, (1972).
	
	
	
	\bibitem{bagmorc} F. Bagarello, G. Morchio, {\em Dynamics of mean field spin models from
		basic results in abstract  differential equations}, J. Stat. Phys.
	{\bf 66}, 849-866 (1992)
	
	\bibitem{bagmaster}  F. Bagarello,  {\em Thermodynamical Limit and Boundary
		Effects in  Long Range Spin Models},  Master thesis at SISSA, (1989)
	

	
	
	


\end{thebibliography}
\end{document}